# Spoken Digit Classification by In-Materio Reservoir Computing with Neuromorphic Atomic Switch Networks

Sam Lilak[1], Walt Woods[2], Kelsey Scharnhorst[1], Christopher Dunham[1], Christof Teuscher[2], Adam Stieg[3,4,*], and James Gimzewski[1,3,,4,5*]

[1]University of California, Los Angeles, Department of Chemistry and Biochemistry, Los Angeles, CA 90095, USA

[2]Portland State University, Department of Electrical and Computer Engineering, Portland, OR 97201, USA

[3]University of California, Los Angeles, California NanoSystems Institute, Los Angeles, CA 90095, USA

[4]WPI Center for Materials Nanoarchitectonics (MANA), National Institute for Materials Science (NIMS), Tsukuba 305-0044, Japan.

[5]Research Center for Neuromorphic AI Hardware, Kyutech, Kitakyushu 8080196, Japan.

**Correspondence:** stieg@cnsi.ucla.edu, gim@chem.ucla.edu

**Keywords:** atomic switch networks, memristive, neuromorphic, reservoir computing

**Abstract**. Atomic Switch Networks (ASN) comprising silver iodide (AgI) junctions, a material previously unexplored as functional memristive elements within highly-interconnected nanowire networks, were employed as a neuromorphic substrate for physical Reservoir Computing (RC). This new class of ASN-based devices has been physically characterized and utilized to classify spoken digit audio data, demonstrating the utility of substrate-based device architectures where intrinsic material properties can be exploited to perform computation in-materio. This work demonstrates high accuracy in the classification of temporally analyzed Free-Spoken Digit Data (FSDD). These results expand upon the class of viable memristive materials available for the production of functional nanowire networks and bolster the utility of ASN-based devices as unique hardware platforms for neuromorphic computing applications involving memory, adaptation and learning.

## 1. Introduction.

Speech recognition is a seminal task in the field of artificial intelligence and natural language processing. Typical algorithmic approaches to speech recognition break apart sections of raw speech data and bin them into hidden Markov models manipulating Markov chains. While effective, these approaches are more computationally intensive than some recently developed neural network models, which may prove a more suitable compute framework for handling increasingly larger data sets[1–3]. Artificial Neural Networks (ANNs) have also been a promising avenue for more efficient speech recognition tasks which offer the benefit of being trained for natural language processing and are believed to be a more suitable candidate for handling the varied complexity of each person's unique voice and accent. Implementation of ANNs in modern computing hardware remains computationally burdensome and often requires access to and utilization of high-performance computing clusters. A suitable hardware architecture for local execution of complex tasks such as natural language processing must be able to process dynamic, temporal data in real-time while remaining energy efficient. Memristive materials have been identified as strong candidate for such applications as they offer an opportunity to alleviate the bus latency between memory and processing elements in traditional von Neumann architectures while also performing in-memory computation with reduced power consumption[4]. The nonlinear character of memristors, resulting from the underlying physics of the material itself, is essential for enabling simultaneous storage of data (memory) and performance of complex tasks with it (processing) through a relatively new technique known as evolution in-materio[5–8].

The growing field of evolution in-materio computing has sought to optimize computational architectures via evolutionary (search) algorithms[6,7]. The materials and architectures employed vary with the desired facet of computation, but ideally these materials are computationally and energetically efficient at employing a litany of machine learning based algorithms. Utilizing a single hardware element capable of exhibiting both memory and processing alleviates the burden of busing information between two separate hardware components, reducing latency in computation[3]. The most robust currently known architecture that combines the aforementioned elements is the mammalian brain, which has been both a foundation and inspiration towards the development of architectures which can efficiently process multi-input, chaotic, and/or time-varying (temporal) datasets.

This work focuses on the class of neuromorphic computing devices known as Atomic Switch Networks (ASN), comprising a highly-interconnected network of memristive nanowire junctions as shown schematically in Figure 1. Ongoing efforts to develop memristive hardware for neuromorphic computing include not only ASNs, but also patterned crossbar arrays, and nanoparticle clusters[9–13]. ASN-based devices provide a physical system with structure and functional dynamics reminiscent of the mammalian brain[14–17] that has previously been employed as a computational material for applications in Reservoir Computing (RC)[18–24]. The atomic switch is a nanoscale electroionic element consisting of a Metal-Insulator-Metal (MIM) junction whose properties can be manipulated via a time-dependent input signal[25–28]. Individual atomic switches have been shown to produce memristive, nonlinear responses, exhibiting both short and long-term memory as well as quantized conductance[12,13,29,30]. These properties render atomic switches and other memristive systems as ideal circuit elements for use within a network architecture that can serve as a dynamic physical reservoir used to solve complex computational tasks, including speech recognition and natural language processing[31,32].

RC provides a framework for computing complex functions using a dynamical system as a 'reservoir'[18,33–35]. The RC framework is ideal for the processing of dynamic, temporal real-time signals and can be used in many of the same situations as recurrent feed-forward neural networks. RC also offers advantages such as fault-tolerance and the capacity for learning[35,36]. Passing a time varying input through a dynamic reservoir produces a higher dimensional representation of the signal through nonlinear transformation, where different points on the reservoir are measured and linearly combined to reproduce an arbitrary output signal as shown in Figure 2. Training is only performed on the linear readout coefficients; the reservoir dynamics themselves are generally considered fixed. Limiting training to the weights between the reservoir and output layer alleviates the need to use gradient-descent based methods, greatly minimizing the associated computational burden.

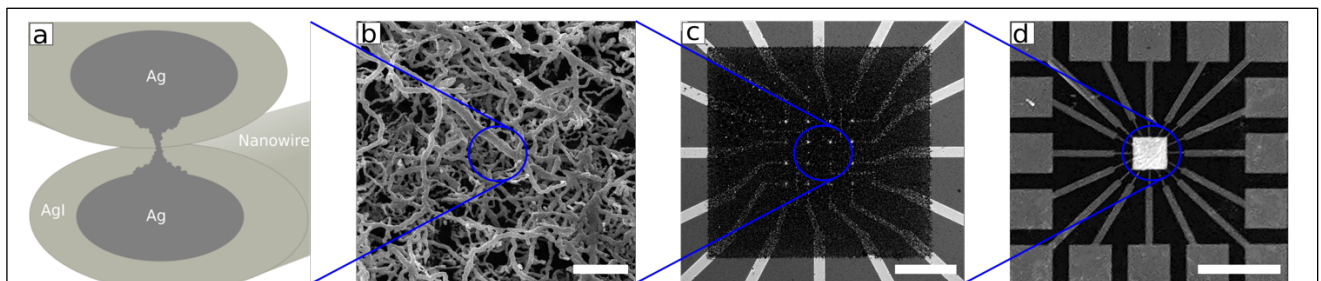

Figure 1. Schematic diagram of an AgI-based ASN device, from nanowire junction to chip (left to right). (a) Ag-AgI-Ag junctions form between overlapping nanowires. Yellow-gray represents AgI. Dark-gray represents Ag. Filament formation occurs as a gapless junction between Ag nanowires. (b) SEM image of the interconnected nanowire (scale bar = 20 um). (c) Optical image of a microelectrode array at center of the ASN device (scale bar = 360 um). (d) Optical image of a complete 16-electrode ASN device (scale bar = 5 mm).

As an alternative to simulation-driven RC, in-materio RC leverages material complexity for computational purposes[37,38]. Whereas early implementations of RC simply utilized a body of a liquid acting as the dynamic reservoir, more recent works harnessed the intrinsic properties of complex physical systems, including ASNs, as the basis for a computation[20,22,24,33,39].

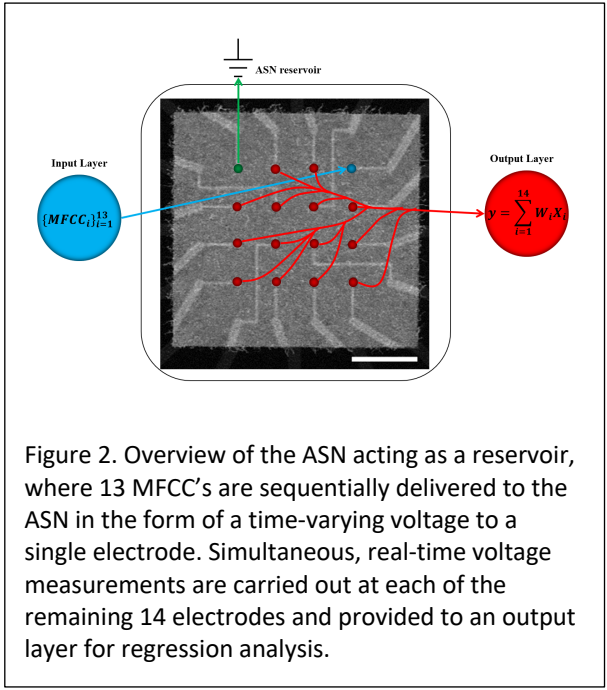

Figure 2. Overview of the ASN acting as a reservoir, where 13 MFCC's are sequentially delivered to the ASN in the form of a time-varying voltage to a single electrode. Simultaneous, real-time voltage measurements are carried out at each of the remaining 14 electrodes and provided to an output layer for regression analysis.

Computational neural models such as the perceptron and support vector machine can also be used as reservoirs; however, long convergence times can be a drawback depending on the task. Material-based reservoirs have the benefit of efficiently performing these tasks in-situ, enabling low-power, on-chip computing[40–42]. This alternative approach offers the opportunity to employ neural networks and machine learning algorithms offline, without the need to access servers, clusters and other high-performance computing infrastructures.

ASNs have been shown to represent uniquely suitable class of materials for implementation of hardware-based RC, namely complex network architectures with the requisite material complexity[15,36,43]. These self-organized systems offer a unique opportunity to produce highly-interconnected memristive networks, where a density of atomic switch junctions of up to $10^9/cm^{30}$ has been previously reported. The fabrication scheme, based on electroless deposition, produces a diverse ensemble of silver nanowires with varying lengths, widths and thereby junction dimensions. This structural diversity in the material substrate imparts a distribution of operational characteristics that improves the capacity to perform non-linear transformations of input signals.

Herein, we report the use of a new memristive material, silver iodide (AgI), as the functional element in the ASN framework[13,44,45]. Silver iodide can be robustly prepared in a brief vapor phase reaction of iodine vapor with silver nanowires at room temperature in contrast to the lengthy formation times at elevated temperatures of previously reported silver sulfides. This promising material provides voltage-controlled resistance in both the bulk and when integrated into crossbar architectures, rendering it suitable as a memristive material for RC applications which require non-linear transformations and quantized conductance states[36]. This work expands the catalog of investigated ASN materials by fabricating and testing AgI for non-linear, temporal computation through the classification of spoken digits.

**2. Methods**

*2.1 Device Fabrication*

The substrate for ASN devices, a multielectrode array enabling spatiotemporal stimulation and monitoring, was fabricated using standard thermally oxidized (500 nm) silicon wafers as the base substrate. A 16-electrode grid of Pt (150 nm) was patterned by photolithography and deposited using a negative photoresist (AZ NLOF 2020) onto a Cr or Ti wetting layer (5 nm). Liftoff was induced overnight in N-methyl-2-pyrrolidone (NMP) at 60°C. Point contact electrodes were prepared using a patterned insulating layer of SU-8 (400 nm) which was soft baked (90°C), exposed to UV, post exposure baked (90°C), developed for 3 min, and hard baked at 180°C for 30 min. An array of copper (300 nm) seed sites with 5 x 5 μm spacing in a grid were patterned onto inner point contact electrodes and deposited onto AZ NLOF 2020 via metal evaporation at 3 nm/s followed by lift-off overnight in NMP (60°C). The resultant device platforms consist of a stack of Si/SiO2/Cr/Pt-

electrodes/SU-8/Cu-posts (Figure S1) and were stored in inert atmosphere until bottom-up silver nanoarchitecture construction[34,46].

This substrate was placed into a 50 mM solution of silver nitrate (AgNO₃) for 30-60 minutes. Silver nanowires formed through an electroless deposition reaction involving the reduction of silver and the oxidation of copper through the following reaction:

$$Cu^0_{(s)} + 2Ag^+_{(aq)} \rightarrow Cu^{+2}_{(aq)} + 2Ag^0_{(s)}$$

The ordered copper posts (5x5 µm) directed a density-controlled formation of interconnected silver nanowires, whereby each ASN exhibited a unique structure determined by the bottom-up fabrication of metal cations. Subsequent silver iodide was formed in a nitrogen purged and sealed glass chamber with the ASN chip suspended over a small iodine pellet. Two different experimental techniques, one under ambient conditions (5 minutes exposure time) and the other with added heat (30 °C, 2-3 minutes exposure time) were employed with both techniques successfully iodizing the silver nanowires.

$$2Ag_{(s)} + I_{2(g)} \rightarrow 2AgI_{(s)}$$

UV-Vis and XPS samples were prepared using transparent silver thin films (20 nm). These films were deposited on glass cover slides via a silver target in a Hummer 6.2 sputter system at 15 mA from Anatech Ltd (Hayward, CA, USA) under an argon vacuum environment (80 mtorr).

*2.2 Material Characterization*

Optical and scanning electron microscopy (SEM) were used to characterize the as-fabricated structure of the nanowire network. SEM images were acquired using the JEOL JSM-7500F. X-Ray photoelectron (XPS) and UV-VIS spectroscopy were employed using transparent Ag thin film substrates with Ag as a control. Absorbance spectra of thin films were collected using the HP 8453 spectrophotometer. XPS spectra were obtained on an AXIS Ultra DLD XPS instrument from Kratos Analytical. The X-ray source was Al K at 1486.6 eV. Survey (1200 eV) and high-resolution scans were integrated over 4 and 16 sweeps, respectively.

*2.3 Electrical Characterization*

Characterization of ASN devices involves the spatially-defined stimulation and monitoring of electrical activity throughout the network in the form of current and voltage traces. All input-output signals were generated/acquired using a purpose-built software package developed in Labview in conjunction with dedicated hardware manufactured by National Instruments. A data acquisition card (DAQ) (model PXIe-6368) was used to deliver input signals routed through a shielded connector box (model SCB-68A) to the ASN device. A source measurement unit (model PXIe-4141) was used to measure current flow through the ASN at user-selected electrodes, where acquired and applied signals were routed using a 16x32 switch matrix terminal block (model TB-2642B). Voltage traces were simultaneously monitored at all 16 electrodes using the DAQ card. All components were housed in a National Instruments chassis (model PXIe-1078) with an embedded controller.

Prior to any FSDD output signals, each ASN was driven through an initialization (activation) process in which the electrodes were sequentially stimulated with 7 Hz triangle waves. This process was repeated with increasing voltages (0.01-1 V) to realize switching patterns within the network. The switch matrix was employed in conjunction with the DAQ to calculate the resistance of every electrode combination prior to and after initialization, where successful activation was characterized by a sharp reduction in the network-wide parallel resistance as compared to the virgin metal system. Current-voltage and voltage-voltage measurements utilized triangle wave outputs from the DAQ card. The FSDD signal outputs were also produced by the DAQ card at selected electrode locations via the switch matrix.

*2.4 Reservoir Computing*

Spoken digit classification was implemented in AgI ASN devices via RC using the FSDD (Figure 3). The task was not performed using raw audio data, but rather using Mel-Frequency Cepstrum Coefficients (MFCCs) of the data, similar to previously reported techniques. Each 8 kHz wave-format sound file from the FSDD was zero-padded up to 1 second of recording length and then converted into MFCCs using the "python_speech_features" Python package. Mel-frequency cepstrum is a short-term power spectrum of the sound waves, using a linear cosine transform of a log power spectrum and is a nonlinear mel scale of frequency that approximates the human auditory response better than standard linear spacing of frequency components.

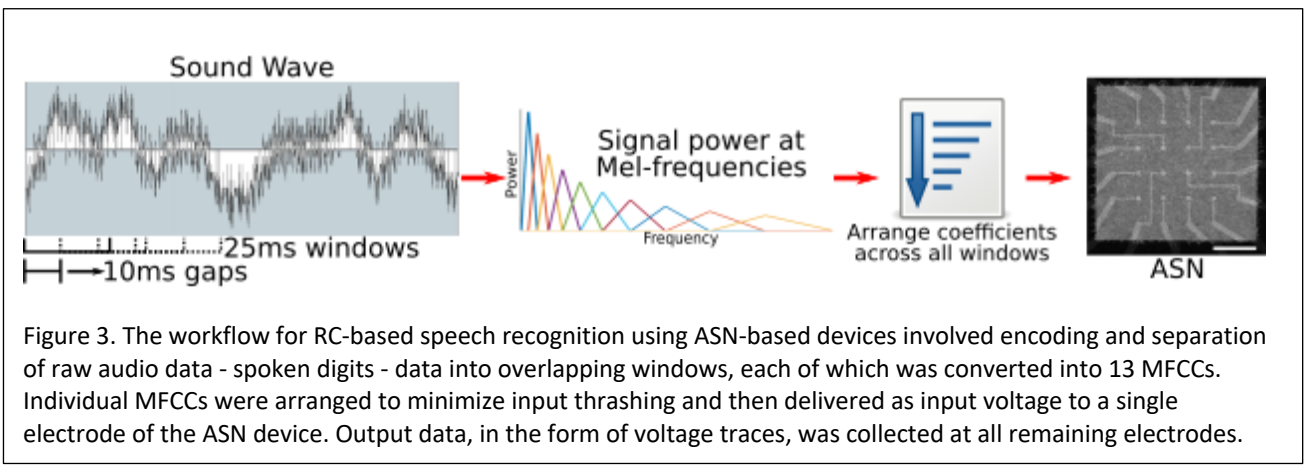

Figure 3. The workflow for RC-based speech recognition using ASN-based devices involved encoding and separation of raw audio data - spoken digits - data into overlapping windows, each of which was converted into 13 MFCCs. Individual MFCCs were arranged to minimize input thrashing and then delivered as input voltage to a single electrode of the ASN device. Output data, in the form of voltage traces, was collected at all remaining electrodes.

Default settings were used, resulting in an array of MFCCs where each 25 ms window of signal was parameterized by 13 MFCCs. Windows were offset by 10 ms, resulting in 1287 total coefficients. To reduce device thrashing, the resulting MFCC array was flattened and fed to the network one at a time. The entire temporal sequence of the lowest-frequency coefficient was passed first, then the next-lowest-frequency coefficient's values, and so on. The resulting 1287 Hz signal was sent to an input electrode, 14 electrodes were measured, and another electrode was grounded. Both the input and 14 read electrodes were recorded at 1 kHz. For RC, the resulting voltage streams were sampled at the end of sub-windows of computation, and the entire collection of sampled recordings was linearly regressed to indicate which digit was spoken (see Figure 4). Twelve unique spoken digit recordings were used, characterized by two speakers, saying three digits, two unique times. The FSDD speakers were 'Jackson' and 'Theo', the digits spoken were zero, one, or two, and the first two instances of each digit were used. As a baseline, regressions were performed on only the input electrode's voltage reading ("input only" mode) as well as on the full electrode suite of the input electrode and the 14 readout electrodes ("reservoir" mode).

## 3. Results and Discussion

### 3.1 Material and Device Characterization

Silver nanowire networks like those shown in Figure 1b were reliably produced based on previously developed protocols. The network functionalization process requires conversion of silver nanowire junctions to silver iodide. The protocol for the formation of silver iodide was validated using UV-Vis and X-ray Photoelectron Spectroscopies (XPS). Figure 4 provides representative visible absorption spectra of as-prepared Ag and AgI thin films.

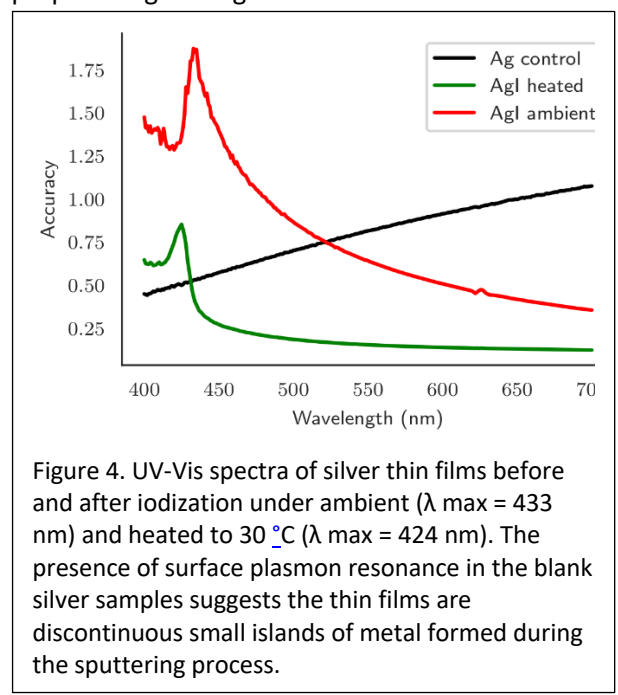

Figure 4. UV-Vis spectra of silver thin films before and after iodization under ambient (λ max = 433 nm) and heated to 30 °C (λ max = 424 nm). The presence of surface plasmon resonance in the blank silver samples suggests the thin films are discontinuous small islands of metal formed during the sputtering process.

Ag thin films prepared by desktop sputtering exhibited a Surface Plasmon Resonance (SPR), suggesting the presence of silver islands within the film[47]. These results are in line with previous reports which have demonstrated that silver exposed to iodine decreases SPR intensity coupled with a buildup of excitons. An absorbance peak around 420 nm has been previously reported and longer exposure to iodine at ambient temperature yielded a red-shifted maximum, which has been associated with the formation of larger AgI particles[47,48].

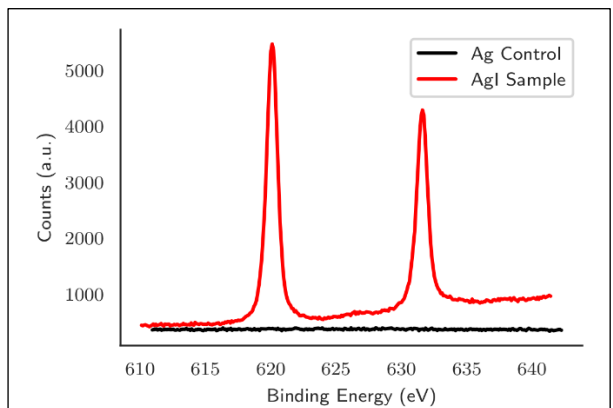

Figure 5. XPS spectra of the iodine 3d $_{5/2}$ and 3d$_{3/2}$ core levels in silver-based ASN devices exposed to (sample, red) and not exposed to (control, black) iodization procedures. The two peaks at 620 and 631 eV correspond to the expected I⁻ bands for I$_{3/d7}$.

XPS results shown in Figure 5 confirmed the presence of characteristic peaks for iodide 3d$_{5/2}$ and 3d$_{3/2}$ core level energies previously reported in metal iodides at binding energies of 620 and 631 electron volts (eV) which are absent in silver control samples[49]. While both functionalization protocols successfully produced AgI, the heated method was used for all ASN devices due to quicker sublimation of solid iodine.

To confirm the viability of AgI networks as a physical substrate for in-materio RC, the spatially distributed nonlinear characteristics of the ASN were examined. Voltage traces acquired at each of the 14 measurement electrodes enabled the analysis of Lissajous plots (V-V) as shown in Figure 6. AgI devices demonstrated distributed nonlinear dynamics throughout the entirety of the nanowire network as a consequence of their highly interconnected nature, where a stable and reproducible nonlinear transformation of the input signal was observed. This capacity for the non-linear transformation of time-varying signals and temporal datasets renders the AgI nanowire network ideal for the performance RC-based speech recognition tasks.

### 3.2 AgI ASN-based Reservoir Computing

AgI nanowire networks were evaluated for their RC potential in spoken digit recognition as shown schematically in Figure 3. To effectively benchmark the value of the nanowire network in the performance of a spoken digit classification task, linear regression was performed in two ways. First, linear regression of the input voltages only - defined as 'Input Only'-was carried out in the absence of the physical reservoir. Second, the full reservoir system – defined as 'Reservoir' - employed regression of both the input signal and all device outputs. Inclusion of the input signal allows the regression to more accurately discern correlations between the transformed output signals and the input itself. FSDD digits encoded as MFCCs and passed to the network as a temporal sequence at 1287 Hz were successfully classified as shown in Figure 7. A sufficient number of training examples were found to stabilize the reservoir's behavior, and evaluating testing data on only a single array of readout coefficients was found to be valid[50].

The target function was regressed by dividing the measured electrode data into $N$ segments and using the last data point from each segment. For the "input only" mode, this means that $N$ = 80 used 80 values in the regression. For the "reservoir" mode, this means that $N$ = 80 used 80 × 15 = 1200 values in the regression. To determine the accuracy at each $N$ value, 12-fold cross-validation was employed using 11 of the audio files as training data and the 12$^{th}$ audio file as testing data. Each file was delivered to the device multiple times on a loop, aggregating far more than twelve tests to compute the accuracy. Nonetheless, there were only 12 unique data streams used. As a result, this problem suffered from significant overfitting, indicatedby the "input only" results decreasing in accuracy as more points were used for the regression. This overfitting manifested as significant noise in the accuracy; $N$ = 100 might give an accuracy as high as 100%, while $N$ = 101 would give an accuracy of

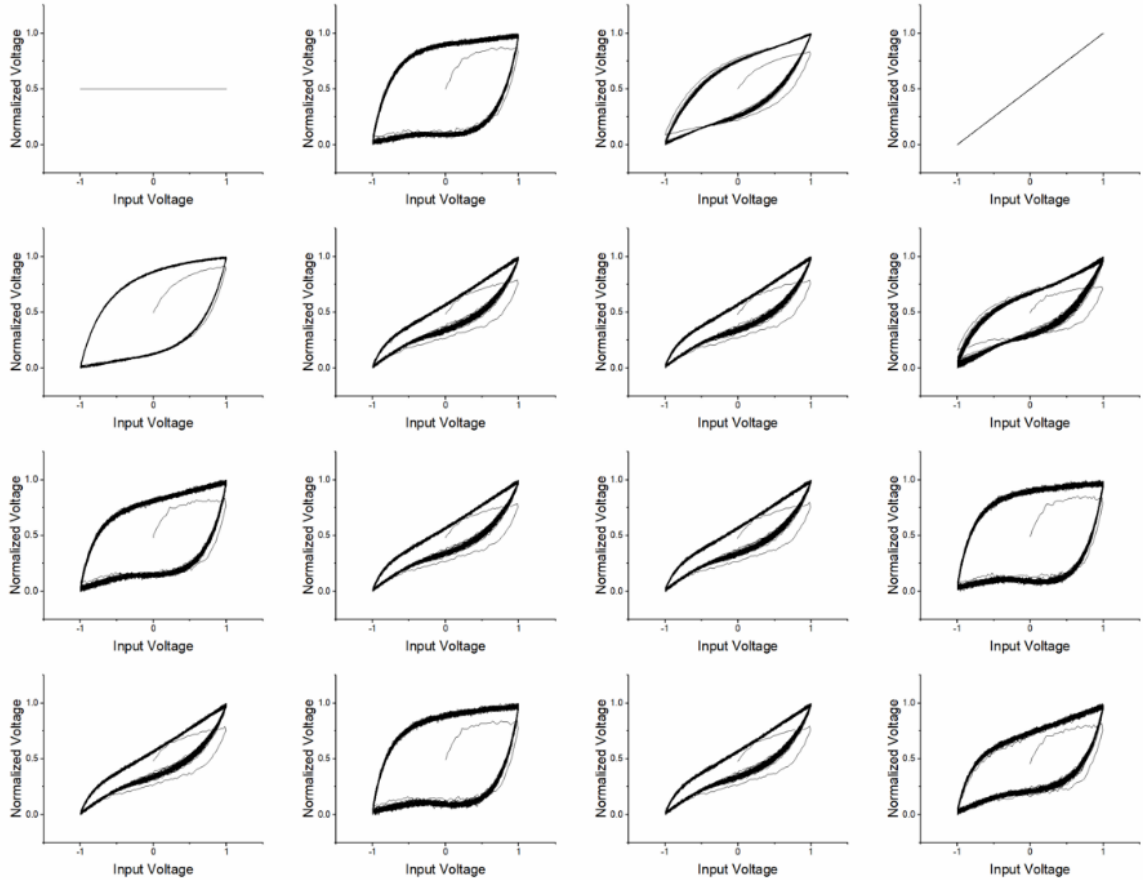

Figure 6. Representative normalized Lissajous plots of all 16 electrodes measured simultaneously using a 7 Hz triangle waveform swept from –1 to +1 V over the course of 23 seconds with the grounded electrode (top left) and input signal (top right) recorded. The network demonstrated a spatially diverse system with reproducible, non-linear behavior distributed throughout the networks.

54 %. To account for this, the space of points N tested was divided into windows of size 25, and the average and standard deviation of accuracy within this window is shown in Figure 7. For instance, the mean and standard deviation shown at $N = 100$ indicate the statistics for $N \in \{88, ..., 112\}$. The ASN reservoir also demonstrated highly accurate results across a wide range of input voltages (0.5-10V), suggesting potential utility of these devices for low-power applications.

These results clearly demonstrate the added stability provided by the ASN reservoir, evidenced by consistent accuracy at higher points of regression in the reservoir. The ASN's robustness and versatility was demonstrated by its capability to discern spoken digits when stimulated by both high and low voltage signals without a significant loss in accuracy. The ASN also provided a moderate benefit in accuracy, even before the input-only lines began overfitting. The lack of overfitting on the reservoir lines could be interpreted as a side-effect of the temporal, non-linear properties of the reservoir. This is corroborated by the fact that the reservoir lines achieved higher accuracy than the input only lines, a phenomenon that could not be achieved without non-linear or temporal behavior. Rather than relying on a stream of individual values, each of which has some noise associated, the reservoir readout mode could rely on 15 such streams. Assuming the noise on each electrode is somewhat independent, averaging these channels could have significantly reduced noise.

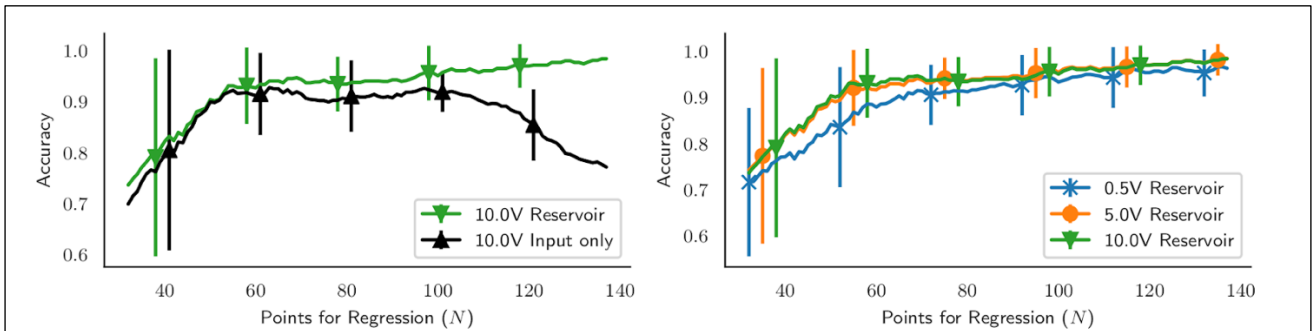

Figure 7. (Left) Performance of the spoken digit classification task using AgI nanowire networks for in-materio RC to tap the temporal sequence of spoken digit MFCCs at N different points and regressing to identify the digit spoken. Mean accuracy and standard deviation clearly shows that the 'Reservoir' readout method avoided overfitting and improved task performance as compared to using the 'Input Only' mode. (Right) The input signal amplitude (voltage) was observed to have minimal. impact on accuracy, indicating the potential for maintaining task performance under low-power operation of AgI ASNs.

## 4. Conclusion

Neuromorphic nanowire networks such as the ASN represent a burgeoning class of material architectures whose dynamical nature makes them uniquely suited to serve as physical substrates for hardware-based, in-materio computing. While the ever-increasing demands for computational capacity and complexity continue to challenge even the most advanced computing architectures, dynamical in-memory compute platforms such as the ASN may provide an alternative solution that is scalable, energy-efficient, adaptive, and capable of processing complex, time-varying data without the need for pre-programming or remote intervention. Expanding the catalog of memristive materials amenable to production of ASN-based devices, and thereby the diversity of network dynamics available for task performance, further increases their potential utility as a platform technology for next-generation computing applications. The new AgI-based ASN devices reported here served as a dynamic, memristive reservoir for the nonlinear transformation of temporal data and demonstrated the capacity to reliably classify spoken digits with high accuracy across a wide range of input voltages. Combined with the relative ease and low cost of the fabrication process, these AgI nanowire networks represent both a new material system that is ripe for future study and an opportunity to further develop the concept of in-materio computing toward real-world applications.

## 5. Conflict of Interest

No conflicts of interest to declare.

## 6. Funding

This work was partially supported by the World Premier International Center for Materials Nanoarchitectonics (MANA) at the National Institute for Materials Science (Tsukuba, Japan). This material is based upon work supported by the National Science Foundation Graduate Research Fellowship under Grant No. (NSF grant number), and Semiconductor Research Corp. under Grant No. (2015209024). The views expressed are those of the author(s) and do not reflect the official policy or position of the Department of Defense or the U.S. Government. Approved for public release, distribution is unlimited.

## 7. Acknowledgements

The authors gratefully acknowledge past group members: Cristina Martin-Olmos, Henry Sillin and Audrius V. Avizienis for their foundation work in chip fabrication, LabView programming and crystal growth, respectively. Physical ASN chips were fabricated in the UCLA Nanofabrication Laboratory at the California Nanosystems Institute (CNSI). Instruments used in this study were operated and maintained by the Molecular Instrumentation Center at the Department of Chemistry and Biochemistry at UCLA.

## 8. Author Contributions

S.L. and W.W. contributed equally to this work. A.S., W.W. and K.S. conceived and designed the experiments. S.L., W.W., K.S. and C.D. performed experiments and analyzed data. All authors discussed the results and contributed to preparation of the manuscript. S.L., W.W., K.S. and A.S. wrote the manuscript. C.T. and J.G reviewed and edited the manuscript.

## 9. Data Availability Statement

The datasets generated from this work are available upon request.